\documentclass[aps,prappl,reprint,longbibliography]{revtex4-1}

\usepackage{graphicx}
\usepackage{dcolumn}
\usepackage{bm}
\usepackage{epsfig}
\usepackage{amsmath}
\usepackage{mathtools}
\usepackage{subfigure}
\usepackage{epstopdf}
\usepackage{textcomp}
\usepackage{mleftright}
\usepackage{url}
\usepackage[colorlinks]{hyperref}

\hypersetup{%
        plainpages=true,
        breaklinks=true,
        hypertexnames=false,
        pageanchor=true,
        colorlinks=true,
        linkcolor={blue},
        citecolor={magenta},
        urlcolor={blue},
        anchorcolor={black}
      }

\hypersetup{%
        plainpages=true,
        breaklinks=true,
        hypertexnames=false,
        pageanchor=true,
        colorlinks=true,
        linkcolor={blue},
        citecolor={magenta},
        urlcolor={blue},
        anchorcolor={black}
      }

\usepackage{mathrsfs,bm,amsthm,amsfonts,xspace,float,bookmark}
\usepackage{textcomp}
\usepackage{tabu}
\usepackage{bbold}
\usepackage{xspace}
\usepackage{float}
\usepackage[inline]{enumitem}

\newcommand{\kb}[1]{\langle#1\rangle}
\newcommand{\unite}[1]{\,\rm#1}

\newcommand{\omga}[1]{\omega_{\rm#1}}
\usepackage[normalem]{ulem}

\newcommand{\be}{\begin{equation}}
\newcommand{\ee}{\end{equation}}
\newcommand{\bea}{\begin{eqnarray}}
\newcommand{\eea}{\end{eqnarray}}
\newcommand{\beq}{\begin{eqnarray}}
\newcommand{\eeq}{\end{eqnarray}}

\usepackage{lineno}
\begin{document}
%\linenumbers

\author{Yong Lu$^{1,2}$}
\email[e-mail:]{kdluyong@outlook.com}

\author{Marina Kudra$^{1}$}
%\author{Simone Gasparinetti$^{1}$}
\author{Timo Hillmann$^{1}$}
\author{Jiaying Yang$^{1,3}$}
\author{Hangxi Li$^{1}$}
\author{Fernando Quijandr\'{\i}a $^{1,4}$}
\author{Per Delsing$^{1}$}

\email[e-mail:]{per.delsing@chalmers.se}

\address{$^1$Microtechnology and Nanoscience, Chalmers University of Technology, SE-412 96, G\"{o}teborg, Sweden
\\$^2$  3.Physikalisches Institut, University of Stuttgart,70569 Stuttgart, Germany
\\$^3$  Ericsson research, Ericsson AB, SE-164 83, Stockholm, Sweden
\\$^4$  Quantum Machines Unit, Okinawa Institute of Science and Technology Graduate University, Onnason, Okinawa 904-0495, Japan
}
%\title{Tuneable Microwave single-photon source based on pulse cancellation with high purity.}
%\title{Characterization and stability of a tunable microwave single-photon source using a superconducting qubit and pulse cancellation}
%\title{Characterisation of quantum efficiency and stability of a tuneable microwave single-photon source based on a superconducting qubit}
\title{Resolving Fock states near the Kerr-free point of a superconducting resonator}
%\title{Resolving Photon Fock states in a nonlinear resonator}

\begin{abstract}
We have designed a tunable nonlinear resonator terminated by a SNAIL (Superconducting
Nonlinear Asymmetric Inductive eLement). Such a device possesses a sweet spot in which the external magnetic flux allows to suppress the Kerr interaction. We have excited photons near this Kerr-free point and characterized the device using a transmon qubit. The excitation spectrum of the qubit allows to observe photon-number-dependent frequency shifts about nine times larger than the qubit linewidth. Our study demonstrates a compact integrated platform for continuous-variable quantum processing that combines large couplings, considerable relaxation times and excellent control over the photon mode structure in the microwave domain.
\end{abstract}

\maketitle

%\section{INTRODUCTION}
\label{sec1}
Encoding quantum information in the infinite Hilbert space of a harmonic oscillator is a promising avenue for quantum computing. Recently, significant progress has been made by using three-dimensional (3D) microwave cavities~\cite{ma2020error,hu2019quantum,reinhold2020error,grimm2020stabilization,gertler2021protecting,gao2018programmable}. Thanks to the strong-dispersive coupling, quantum states such as cat states~\cite{vlastakis2013deterministically,wang2016schrodinger}, GKP states~\cite{campagne2020quantum} and the cubic phase state~\cite{kudra2021robust} have been engineered. However, currently, the scalability and connectivity is difficult, limited by the size of the cavity.  Another simpler method is to use coplanar microwave resonators where resonators and qubits can be fabricated together in a single chip~\cite{schuster2007resolving,wang2008measurement,hofheinz2008generation}. The drawback is the shorter relaxation time of coplanar resonators compared to 3D cavities.

Both  2D and 3D microwave resonators as well as acoustic resonators \cite{chu2018creation,andersson2019non} host linear modes. Therefore, an ancillary qubit is customarily used to introduce a nonlinearity for state preparation and operation. However, the limited coherence of the ancillary qubit and the imperfect operations on it will decrease the fidelity of the actual states~\cite{cavity2015heeres,heeres2017implementing,kudra2021robust}. To avoid operations on ancillary qubits, a Superconducting QUantum Interference Device (SQUID) can be used to terminate a coplanar resonator. This not only provides the tunability of the mode frequency by changing the external magnetic flux through the loop~\cite{wallquist2006selective,mahashabde2020fast,kennedy2019tunable,palacios2008tunable,vissers2015frequency,sandberg2008tuning}, it also induces a sufficiently nonlinearity. Using this nonlinearity, experiments in waveguide quantum electrodynamics have demonstrated the generation of entangled microwave photons by the parametrical pumping of a symmetrical SQUID loop~\cite{schneider2020observation,sandbo2018generating}. Non-Gaussian states, regarded as a resource for quantum computing, have also been realized~\cite{chang2020observation,agust2020tripartite,quantum2019wang}. Therefore, state preparation and operations can be implemented in a nonlinear resonator, even without ancillary qubits. However, in those experiments, the generated states can not be stored for a long time since the resonators are directly coupled to the waveguides.

Theoretically, it has been shown that pulsed operations on a novel tunable nonlinear resonator can be used to achieve a universal gate set for continuous-variable (CV) quantum computation~\cite{timo2020universal}. The proposed device is similar to a parametric amplifier where the Josephson junction or SQUID loop is replaced by an asymmetric Josephson device known as the SNAIL (Superconducting Nonlinear Asymmetric Inductive eLement)~\cite{frattini2018optimizing,sivak2019kerr}, is applied. Using a SNAIL or an asymmetrically
threaded SQUID loop~\cite{lescanne2020exponential,miano2022frequency}, it is possible to realize three-wave mixing free of residual Kerr interactions by biasing the
element at a certain external magnetic flux. In this work, we refer to this flux sweet spot as the Kerr-free point.

\begin{figure*}[t!]
\includegraphics[width=1\linewidth]{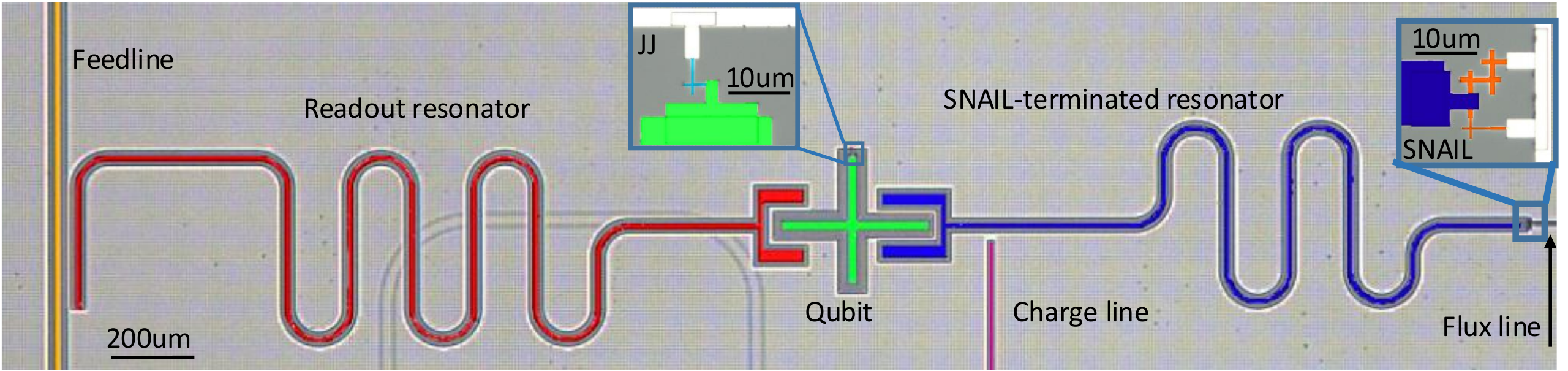}
\caption{\textsf{{\bf{\textsf{A scanning optical micrograph of the measured sample}}.}
A superconducting qubit with a cross-shaped island (green) and a single Josephson junction (JJ, light blue in the middle inset), capacitively coupled to both a coplanar read-out resonator (red) and the nonlinear resonator (blue). The nonlinear resonator is formed by a linear coplanar resonator terminated by a superconducting nonlinear asymmetric inductive element (SNAIL) that has three big Josephson junctions and one small junction (orange in right inset). The charge line (pink) is used to operate the qubit and drive the nonlinear resonator. The qubit state read-out is implemented through the transmission of the feedline. The on-chip flux line is not used in this work, instead, a superconducting coil on the top of the chip [not shown] is used to generate the external magnect flux  $\Phi_{\rm{ext}}$ through the SNAIL. See more fabrication details and the measurement setup for this device in Supplementary.
}
}
\label{chip}
\end{figure*}

\begin{figure}[tbph]
\includegraphics[width=1.0\linewidth]{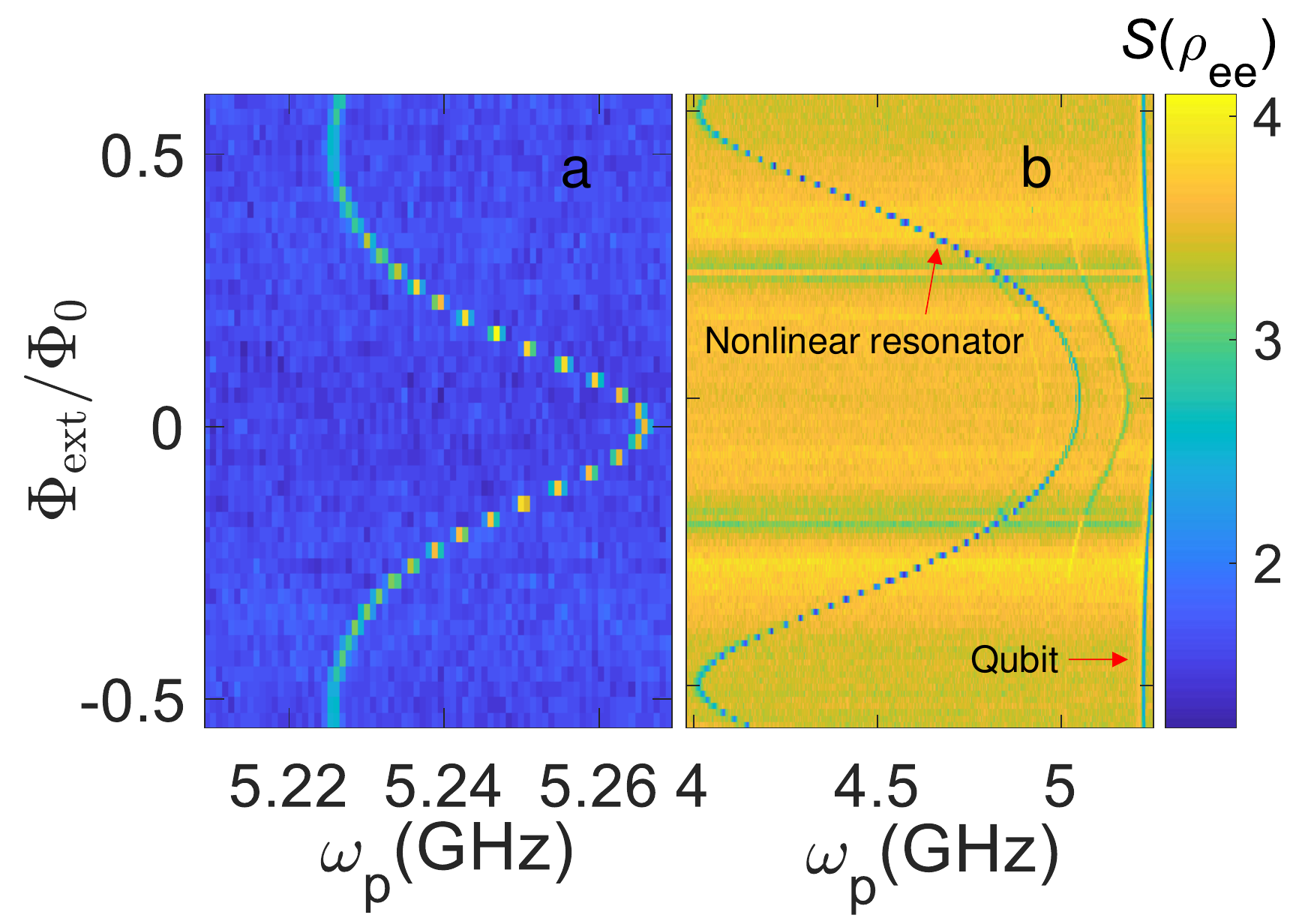}
\caption{\textsf{{\bf{\textsf{Spectroscopy for the qubit frequency and the SNAIL-terminated resonator frequency}}.}
{\bf{a}}, The qubit frequency vs. the external magnetic flux through the SNAIL.
{\bf{b}}, The SNAIL-terminated resonator frequency vs. the external magnetic flux through the SNAIL. The nonlinear resonator is dispersively coupled to the qubit as shown in Fig.~\ref{chip}. $S(\rho_{\rm{ee}})\propto \rho_{\rm{ee}}$ is the measured signal from the read-out pulse of the qubit.
$\omga{p}$ is the probe frequency.
}
}
\label{snailandqubit}
\end{figure}

Therefore, it is meaningful to investigate a SNAIL-terminated resonator in circuit quantum electrodynamics (cQED) where the nonlinear resonator, decoupled to the waveguide, can be used for state preparation and storage.
Strong dispersive coupling between oscillators and qubits  was achieved a decade ago for photons~\cite{schuster2007resolving} and recently for phonons ~\cite{sletten2019resolving,arrangoiz2019resolving,von2021parity}, in linear resonators. Here, we  observed the very well-resolved photon-number splitting up to the $9^{\rm{th}}$-photon Fock state in a SNAIL-terminated resonator coupled to a qubit. Our nonlinear resonator has a considerable relaxation time up to $T_1=20\unite{\mu s}$ under a few-photon drive, limited by (Two-Level Systems) TLSs. Our study opens the door to implement, operate and store quantum states on this scalable platform in the future. Moreover, compared to a linear resonator, our resonator has a non-negligible dephasing rate from its high sensivity to the magnetic flux noise due to the SNAIL loop.

%\section{Results}

\begin{table}
\caption{SNAIL-terminated resonator parameters at zero magnetic flux and near the Kerr-free point. The values are extracted from a transmission coefficient measurement on a SNAIL-terminated resonator coupled to a transmission line. We have $\gamma_{\rm{s}}=1/T_{\rm{s}}=\omga{s}/Q_{\rm{s}}$ where $\gamma_{\rm{s}}$, $T_{\rm{s}}$, $\omga{s}$ and $Q_{\rm{s}}$ are the resonator intrinsic decoherence rate, coherence time, frequency and the internal Q value, respectively. The intrinsic decoherence rate is given by $\gamma_{\rm{s}}=\frac{\Gamma_1}{2}+\Gamma_{\phi}$ where the intrinsic relaxation rate $\Gamma_1=\frac{1}{T_1}$ with the lifetime $T_1$ and the pure dephasing rate $\Gamma_{\phi}=\frac{1}{T_{\phi}}$ with the pure dephasing time $T_{\phi}$.} \label{tab:snail}
  \centering
\begin{tabular*}{\columnwidth}{  @{\extracolsep{\fill}} c c c c c c  @{} }
\hline
\hline
  $\Phi_{\rm{ext}}/\Phi_0$ & $\omga{s}/2\pi$ & $Q_{\rm{s}}$&$\gamma_{\rm{s}}/2\pi$&$T_{\rm{s}}$\\
   ~&~$\rm{GHz}$&~&~$\rm{kHz}$&~$\mu$s\\
\hline
 0 & 5.14&$2.23\times10^5$&23&6.92\\
 0.386 & 4.31&$3.86\times10^4$&112&1.42\\
\hline
\hline
\end{tabular*}

\end{table}

{\bf{Characterization of the SNAIL-terminated resonator.}} In order to characterize the parameters for the SNAIL-terminated resonator directly, we first fabricated a SNAIL-terminated $\lambda/2$ resonator capacitively coupled to a coplanar transmission line [not shown], which is the same as the nonlinear resonator in Fig.~\ref{chip}. By measuring the transmission coefficient through a vector network analyzer similar to measuring conventional resonators~\cite{verjauw2021investigation,calusine2018analysis,kowsari2021fabrication,probst2015efficient}, at the 10\unite{mK} stage of a dilution refrigerator, we can extract the nonlinear resonator frequency at different external magnetic fluxes [see Supplementary].  For our SNAIL-terminated resonator, the inductive energy can be written as~\cite{timo2020universal,frattini2018optimizing,sivak2019kerr}
 \be
U_{\rm{SNAIL}}(\phi)=-\beta E_{\rm{J}}\cos(\phi)-3E_{\rm{J}}\cos\big(\frac{\phi_{\rm{ext}}-\phi}{3}\big),
\ee
where $\beta$ is the ratio of the Josephson energies of the small and the big junctions of the SNAIL, $\phi$ is the superconducting phase across the small junction, $\phi_{\rm{ext}}=2\pi\Phi_{\rm{ext}}/\Phi_0$ is the reduced external magnetic flux, and $E_{\rm{J}}$, related to the Josephson inductance $L_{\rm{J}}$,  is the Josephson energy of the big junctions in the SNAIL. Upon quantization, the Hamiltonian for the SNAIL-terminated resonator becomes
\be
H_{\rm{s}}=\hbar\omega_{\rm{s}}+g_3(a+a^\dag)^3+g_4(a+a^\dag)^4,
\label{Hsl}
\ee
where $g_3$ ($g_4$) is the couplings for the three (four)-wave mixing coupling strength, and $\omega_s$ is the resonator frequency which follows the relation~\cite{timo2020universal}
 \be
\omega_{\rm{s}}\tan{\big(\frac{\pi}{2}\frac{\omega_{\rm{s}}}{\omega_{\rm{r}0}}\big)}=\frac{Z_cc_2}{3L_J},
\label{omega}
\ee
where $\omega_{\rm{r}0}$ describes the bare resonance frequency of the resonator without the SNAIL, $Z_c=58.7\,\Omega$ for the characteristic impedance of the resonator, and $c_2$ is a numerically determinable coefficient for the linear coupling whose specific value depends on $\beta$ and $\phi_{\rm{ext}}$ in our case.

\begin{figure}[t!]
\includegraphics[width=1\linewidth]{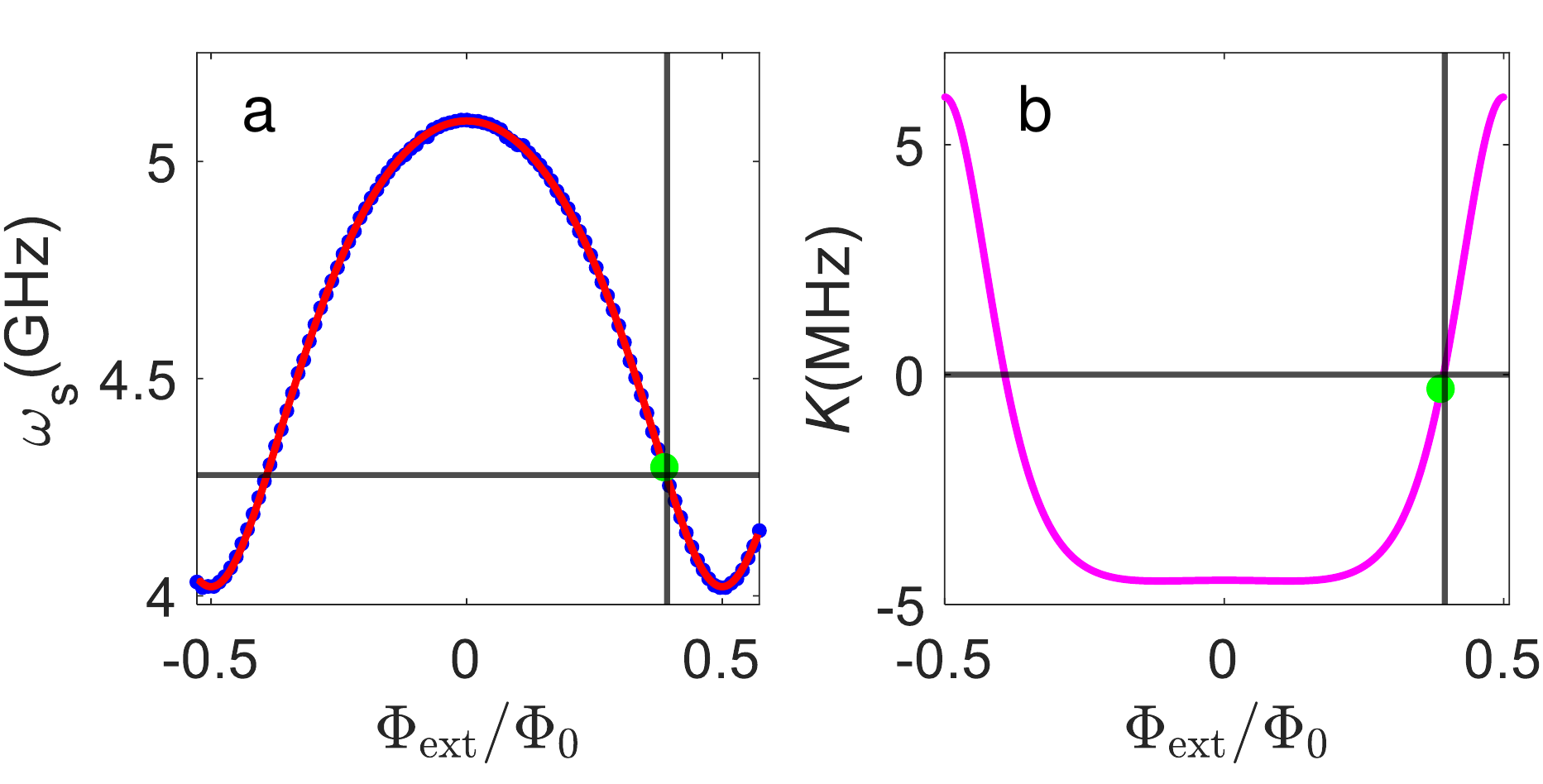}
\caption{\textsf{{\bf{\textsf{Frequency and Kerr strength for SNAIL-terminated resonator coupled to a qubit}}.}
{\bf{a}}, The frequency of the nonlinear resonator vs. the external magnetic flux through the SNAIL. Blue dots (red curve) are (is) the experimental (fitting) data. By fitting the data to Eq.~\ref{omega} [red curve], we obtain $\beta=0.9993\pm0.0005$, well matched with the resistance ratio of the big junction over the small junction 0.104, measured at room temperature, $L_{\rm{J}}=629\pm8\unite{pH}$ and $\omega_{\rm{s}}=8.87\pm0.07\unite{GHz}$. The crossed point from two blacklines indicate the corresponding resonator frequency at the Kerr-free point.
{\bf{b}}, The Kerr-frequency shift per photon, $\it{K}$, vs. the external magnetic flux through the SNAIL calculated from the parameters in {\bf{a}}. The crossed point from two black lines indicates that the Kerr coefficient is zero at $\Phi_{\rm{ext}}=0.392~\Phi_{0}$. The green dot at $\Phi_{\rm{ext}}=0.386~\Phi_{0}$ is the operating flux point in our work.
}
}
\label{snail}
\end{figure}

By sending a very weak probe with the average photon number in the resonator from the probe much less than one, we measure the corresponding transmission coefficient to extract the internal $Q_{\rm{s}}$ values at $\Phi_{\rm{ext}}=0$ and $\Phi_{\rm{ext}}=0.386~\Phi_0$ close to the Kerr-free point, where the magnetic flux is calibrated using the flux period $\Phi_{\rm{0}}$ of the SNAIL-terminated resonator frequency. As shown in Table.~\ref{tab:snail}, we find that the coherence time in both cases is above one microsecond, especially, $T_{\rm{s}}\approx 7\mu s$ at $\Phi_{\rm{ext}}=0$. When we tune the resonator to $\Phi_{\rm{ext}}=0.386~\Phi_0$t, the coherence time is reduced by a factor of five due to the pure dephasing from the magnetic flux noise through the SNAIL.

{\bf{A SNAIL-terminated resonator coupled to a qubit.}} In order to verify the nonlinear resonator performance in cQED in Fig.~\ref{chip},  the nonlinear resonator is dispersively coupled to a fixed frequency superconducting qubit with the bare qubit frequency $\frac{\omega_{\rm{q0}}}{2\pi}\approx 5.222\unite{GHz}$. The effective Hamiltonian that contains the leading order corrections to the rotating wave approximation for describing the coupled SNAIL-terminated resonator and qubit  in the dispersive regime ($g_0 \ll \Delta_0$)~\cite{noguchi2020fast} is
\beq
\frac{H_{\rm{eff}}}{\hbar}\approx\omega_{\rm{c}}a^\dag a+K {a^{\dag2}}a^2+\frac{\omega_{\rm{q}}}{2}b^\dag b-\frac{\chi_0}{2}a^\dag ab^\dag  - \frac{\alpha_q}{2} b^{\dagger 2} b^2,\nonumber\\
\label{Hamitonian}
\eeq
 where the bare SNAIL-terminated resonator is dispersively shifted to be $\omga{c}=\omega_{\rm{s}}-\frac{g_0^2}{\Delta_0}$, and the qubit frequency for the qubit mode $b$ is $\omga{q}=\omga{q0}+\frac{g_0^2}{\Delta_0}$ with the bare qubit frequency $\omga{q0}$ and the Stark shift $\frac{g_0^2}{\Delta_0}$. $g_0$ is the coupling strength between the nonlinear resonator and the qubit, and $\Delta_0=\omga{q0}-\omga{s}$ is the detuning between the resonator and the bare qubit frequencies. The dispersive shift is $\chi_0=\frac{g_0^2\alpha_{\rm{q}}}{\Delta_0(\Delta_0-\alpha_{\rm{q}})}$ with the qubit anharmonicity $\alpha_{\rm{q}}\approx 450\unite{MHz}$.
 Furthermore, near the Kerr-free point where the Kerr nonlinearity (strength $K$) is cancelled in the leading order, the residual Kerr nonlinearity is due to the interplay of four- and three-wave mixing processes~\cite{timo2020universal} as well as the qubit induced nonlinearity.

\begin{figure}[t!]
\includegraphics[width=1\linewidth]{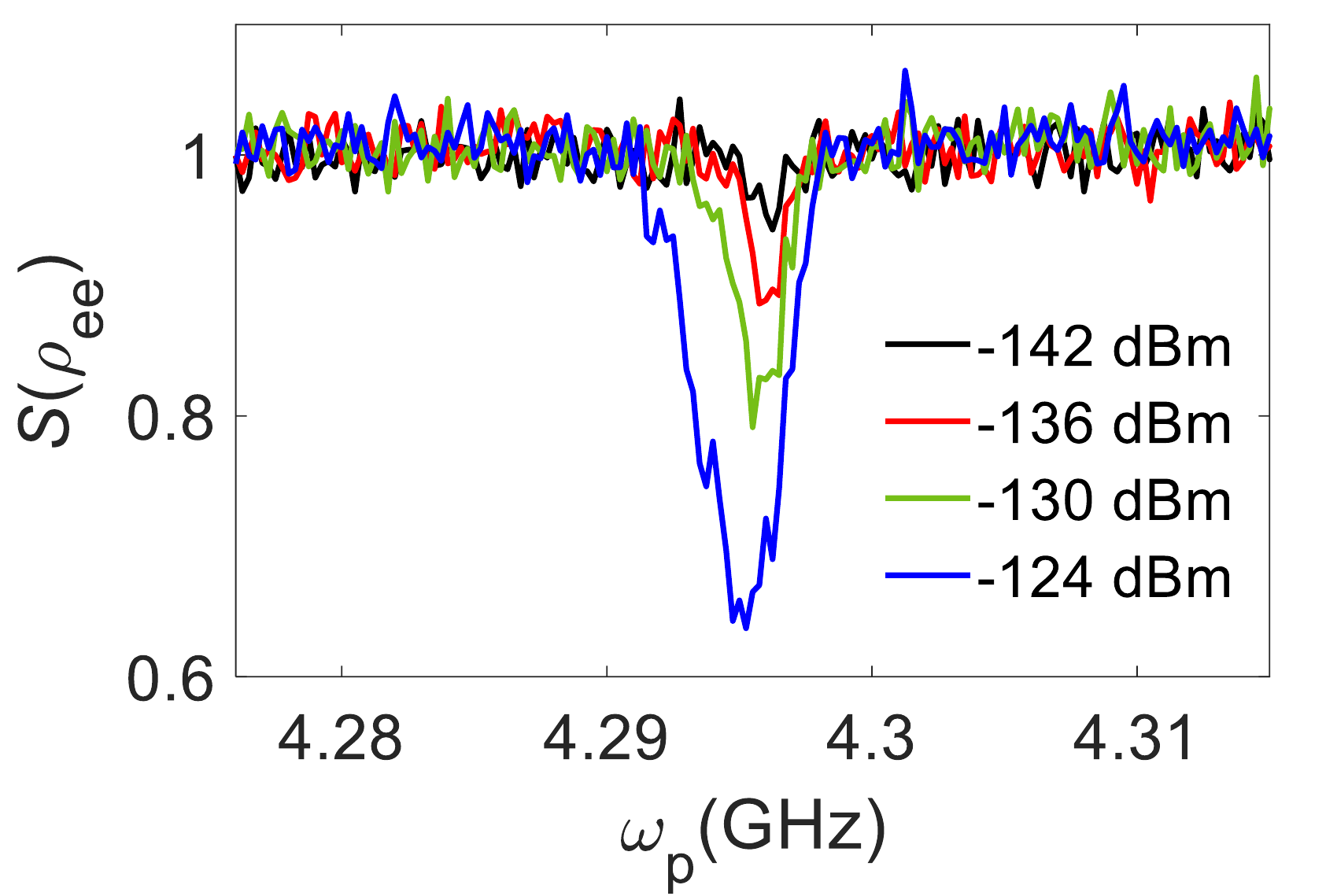}
\caption{\textsf{{\bf{\textsf{The spectroscopy of the nonlinear resonator near the Kerr-free point with different input powers}}.}
The signal $S(\rho_{\rm{ee}})$ is from the readout pulse, depending on the qubit population $\rho_{\rm{ee}}$ where $\rho_{\rm{ee}}$ is determined by the nonlinear resonator photon population. $\omega_{\rm{p}}$ is the frequency of the continuous probe where we vary the probe power to the sample from $-142\unite{dBm}$ to $-124\unite{dBm}$. The dip indicates the resonator frequency. The full linewidth of the dip is also reduced from about 10\unite{MHz} to 3\unite{MHz}. Note that the dip linewidth is almost three orders of magnitude larger than the resonator linewidth. At the lowest pump power, we find the qubit-dressed resonator frequency is $\omga{c}/2\pi\approx 4.296\unite{GHz}$.
}
}
\label{snailfine}
\end{figure}
\begin{figure*}[t!]
\centering
\includegraphics[width=0.9\linewidth]{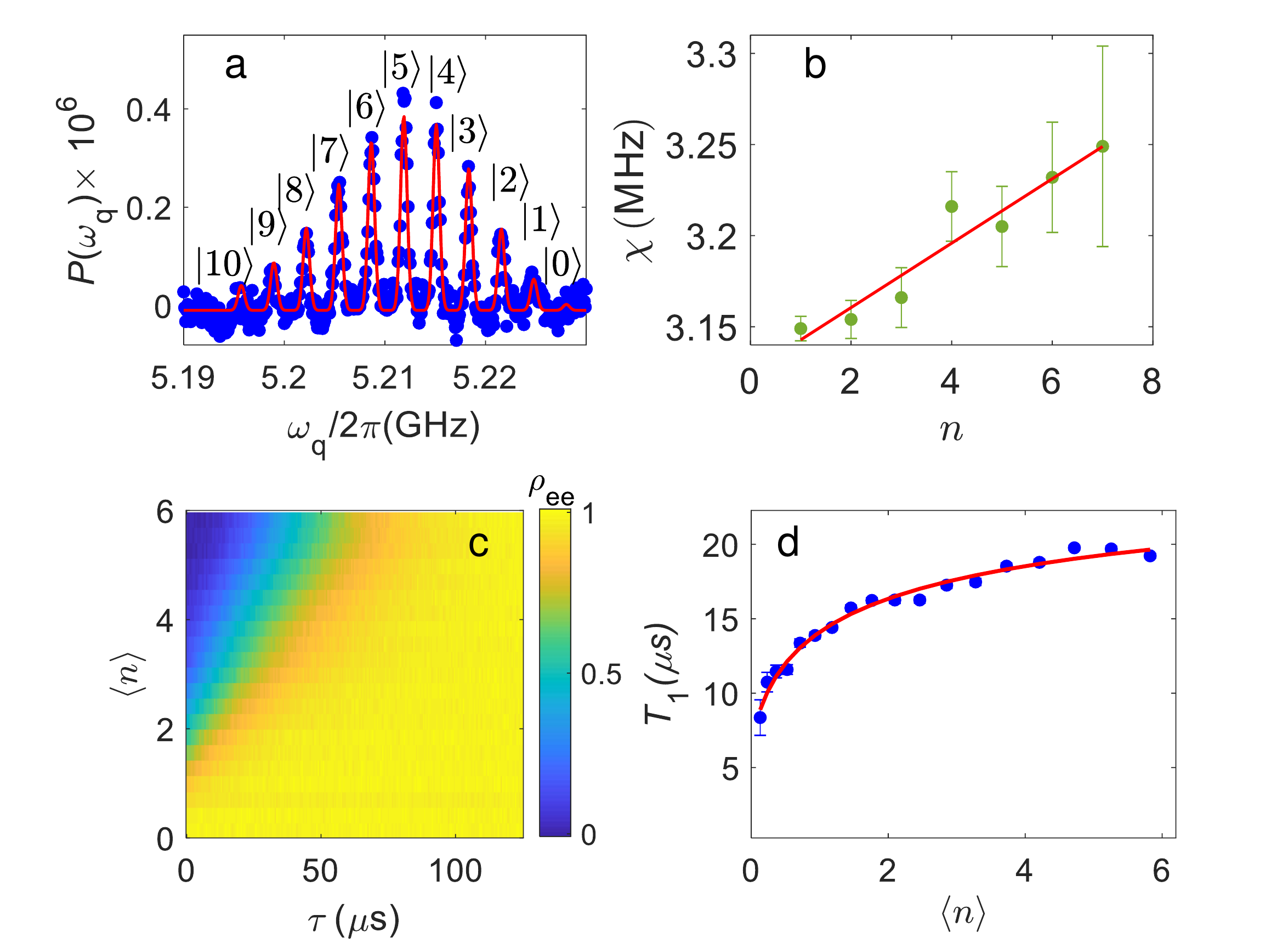}
\caption{\textsf{{\bf{\textsf{Photon-number splitting and lifetime of the SNAIL-terminated resonator near the Kerr-free point.}}}
{\bf{a}}, the qubit-frequency distribution $P(\omega_{\rm{q}})=P(\omega_{\rm{q}})\times 10^6$ with the average photon number $|\alpha|^2\approx 6$, of the coherent displacement in the nonlinear resonator. The envelope of the distribution matches well with a Poisson distribution with the corresponding coherent-state amplitude  $|\alpha|\approx 2.4$ [red solid line].
%{\bf{b}}, Calibration of the coherent displacement $\alpha$ with the voltage output $A$, of the arbitrary-wavefunction generator at the room-temperature.
{\bf{b}}, Dispersive shift $\chi$ over different photon numbers $n$ inside the nonlinear resonator. The data is fitted to the equation $\chi=\chi_0+\chi'(n-1)/2$ with $\chi_0/2\pi=3.143\pm0.021\unite{MHz}$ and the higher order correction $\chi'/2\pi=35\pm11\unite{kHz}$.
{\bf{c}}, the qubit population $\rho_{\rm{ee}}$ depending on the photon number $\kb{n}=|\alpha|^2$ in the nonlinear resonator. $\tau$ is the waiting time after the short displacement before the qubit read-out.
%{\bf{b}}, The qubit population $\rho_{\rm{ee}}$ vs. the waiting time $\tau$ with $\kb{n}\approx6$.
{\bf{d}}, The relaxation time $T_1$ of the nonlinear resonator vs. different photon number $\kb{n}$ inside the resonator. The error bars are for the standard deviation.
}
}
\label{splitting}
\end{figure*}
To experimentally find the qubit frequency, we sweep the frequency of a qubit-excitation pulse with the pulse length $\tau_{\rm{p}}=1\unite{\mu s}$ (spectroscopy pulse, see pulse generation in Supplementary). When the nonlinear resonator frequency is changed by the external flux $\Phi_{\rm{ext}}$, the values of $\Delta_0$ and the qubit frequency are also changed. In Fig.~\ref{snailandqubit}a, at a fixed $\Phi_{\rm{ext}}$, we sweep the pulse frequency, and when the pulse is on resonance with the qubit, the qubit is excited. Then, we measure the qubit state by sending a readout pulse to obtain $S(\rho_{\rm{ee}})\propto \rho_{\rm{ee}}$, which is manifested as bright dots in Fig.~\ref{snailandqubit}a. After finding the qubit frequency and calibrating the corresponding $\pi$-pulses at different values of $\Phi_{\rm{ext}}$, we send a continuous coherent drive to the nonlinear resonator meanwhile driving the qubit on resonance. However, when the drive is on resonance with the nonlinear resonator, the resonator is populated, leading to a change of the qubit frequency due to the dispersive interaction. Thus, the $\pi$-pulse will not excite the qubit anymore, resulting in a smaller signal $S(\rho_{\rm{ee}})$ [Fig.~\ref{snailandqubit}b].

With the values of $\omga{q0}$ and the dressed-qubit frequencies [Fig.~\ref{snailandqubit}a], we can obtain the values of the Stark shift at different external magnetic fluxes. Therefore, we can compensate the Stark shift from the qubit onto the SNAIL-terminated resonator to obtain the bare resonator frequency $\omga{s}$ [blue dots in Fig.~\ref{snail}a  from Fig.~\ref{snailandqubit}b]. With the parameters extracted from Fig.~\ref{snail}a, we can calculate the Kerr coefficient as a function of external magnetic flux [pink curve in Fig.~\ref{snail}b, see Supplementary]. At the device operation point $\Phi_{\rm{ext}}=0.386\unite{\Phi_0}$ [green dot in Fig.~\ref{snail}b], close to the Kerr-free point $\Phi_{\rm{ext}}=0.386~\Phi_{0}$, we deduce a residual Kerr at the operation point of $K=-0.31\pm 0.04\unite{MHz}$ with $g_3/2\pi= -11.6\pm0.4\unite{MHz}$, and $g_4/2\pi=- 0.128\pm0.004\unite{MHz}$.

Due to the dispersive coupling to the qubit, the qubit can be used as a very efficient probe to determine the frequency of the SNAIL-terminated resonator. In our case, the linewidth of the nonlinear resonator is only a few kHz. Thus, it will be extremely difficult to find the resonator frequency based on the reflection coefficient measurement~\cite{yong2021propagating,lu2021characterizing,lu2021characterizing,lu2021nonequilibrium,lin2020deterministic} through the charge line. However, since our qubit is coupled to the resonator with a dispersive shift up to a few MHz as shown in Fig.~\ref{splitting}c, a few photons inside the resonator will shift the qubit frequency significantly. Consequently, even though the frequency detuning between the pump frequency and the resonator frequency is much larger than the linewidth of the resonator, as soon as the intensity of the pump is strong enough to inject a few photons inside the resonator, we can perform qubit spectroscopy to roughly find the resonator frequency. Afterwards, we can find the resonator frequency more accurately by decreasing the pump intensity [Fig.~\ref{snailfine}].

{\bf{Photon-number splitting near the Kerr-free point.}} Once we have determined the nonlinear resonator frequency, we perform a pump-probe measurement consisting of a short pulse (50\unite{ns}) to excite the resonator followed by a long qubit excitation pulse with $\tau_{\rm{p}}=2\unite{\mu s}$ and frequency $\omga{q}$, along with a read-out pulse at the end to infer the qubit excited-state population. Due to the weak hybridization between the qubit and the nonlinear resonator near the Kerr-free point, the short coherent pulse drives the resonator into an approximately coherent state with amplitude $|\alpha|$. We observed the qubit frequency splitting with the Fock states up to 9 photons where each peak has a Gaussian shape due to the quantum fluctuation of the photon number inside the resonator~\cite{gambetta2006qubit} [Fig.~\ref{splitting}a]. The separation of each peak is about 3\unite{MHz}, ten times larger than the qubit linewidth $\frac{\gamma_{\rm{q}}}{2\pi}\approx 280\unite{kHz}$, leading to well-resolved peaks. The pulse length satisfies $\tau_{\rm{p}}\gg 1/\gamma$, resulting in a good frequency resolution.

From the peak difference of the qubit spectroscopy [see Supplementary], we can extract the dispersive shift which increases with the photon number [Fig.~\ref{splitting}b]. According to the dispersive shift $\chi_0$ and the qubit anharmonicity $\alpha_{\rm{q}}$, we can obtain the coupling strength $g_0/2\pi\approx 53\unite{MHz}$. Moreover, near the Kerr-free point, the self-Kerr $K_{\rm{sq}}$ of the SNAIL-resonator induced by the qubit-resonator coupling  \cite{timo2022cubic,noguchi2020fast} is $K_{\rm{sq}}=-24g_4g_0^2/\Delta_0^2\approx-2\pi\times9\unite{kHz}$, which is more than three hundred times smaller than the dispersive shift and more than 10 times smaller than $\gamma_{\rm{s}}$. Thus, the residual Kerr coefficient from the qubit has a small influence on the states stored in our device, which can be also completely suppressed by slightly shifting the external flux $\Phi_{\rm{ext}}$.

{\bf{Lifetime of a single photon in the nonlinear resonator near the Kerr-free point.}} Even though a statistic study can be done by measuring multiple resonators coupled to waveguides to infer the average lifetime, it is still important to study the relaxation time directly on a specific device in cQED since in reality the device performance varies over different devices due to the imperfect fabrication process.

Here, for the device shown in Fig.~\ref{chip}, utilizing the dispersive coupling to the qubit, we can determine the energy lifetime of the nonlinear resonator $T_1$ at the single-photon level. Again, we send a 50\unite{ns} pulse to displace the resonator with the initial amplitude $|\alpha(\tau=0)|$, and after a time delay $\tau$, we apply a conditional $\pi$-pulse where the pulse is on resonance with the qubit frequency when the resonator is empty. Therefore, the qubit-excitation population depends on the photon number $\kb{n}$. In Fig.~\ref{splitting}c at $\tau\sim 0$, when $\kb{n}<1$, the qubit frequency is still mainly centered at $\omga{q}/2\pi=5.228\unite{GHz}$, leading to the qubit population $\rho_{\rm{ee}}\approx 1$. However when $\kb{n}>1$, the qubit frequency starts to split, leading to a decreased $\rho_{\rm{ee}}$ where the population goes to zero smoothly and reaches $\rho_{\rm{ee}}\approx0$ with $\kb{n}=5.8$. However, with increasing $\tau$, the coherent field in the resonator decreases as $\alpha(\tau)=\alpha(0){\rm{exp}}(-\tau/(2T_1))$ and the qubit frequency moves back. Thus, the conditional pulse can excite the qubit again.

Two-level systems (TLSs) are investigated extensively using conventional coplanar resonators~\cite{woods2019determining,calusine2018analysis,wenner2011surface,burnett2014evidence,de2020two,brehm2017transmission,mcrae2020materials} along waveguides, where the photon number inside the resonator can be only roughly estimated according to the attenuation of the setup, and there is no TLS study in cavity quantum electrodynamics. Here, thanks to the accurate calibration of the photon number inside the nonlinear resonator, it is possible to directly study the TLS effect on our nonlinear resonator. The value of $T_1$ in Fig.~\ref{splitting}d, extracted from the data in Fig.~\ref{splitting}c [see Supplementary], increases from $8\unite{\mu s}$ to $20\unite{\mu s}$, which can be explained by saturation of two-level systems (TLS). According to the TLS model~\cite{mcrae2020materials,burnett2014evidence,gao2008physics}, the resonator internal $Q_1$ is given by
\beq
\frac{1}{Q_1}=\frac{1}{\omga{c}T_1}=F\delta_{\rm{TLS}}\frac{\tanh{\big(\frac{\hbar\omga{c}}{2k_{\rm{B}}T_{\rm{res}}}\big)}}{\sqrt{1+\frac{\kb{n}}{n_{\rm{c}}}}}+\delta_{\rm{other}},
\label{TLS}
\eeq
where $F$ is the filling factor describing the ratio of electrical field in the TLS host volume to the total volume. $\delta_{\rm{TLS}}$ is the TLS loss tangent of the dielectric hosting the TLSs, $n_{\rm{c}}$ is the critical photon number within the resonator to saturate one TLS. $\delta_{\rm{other}}$ is the contribution from non-TLS loss mechanisms. A fit to Eq.~(\ref{TLS}) gives $F\delta_{\rm{TLS}}\approx 4.5\times 10^{-6}$, $\delta_{\rm{other}}\approx 1.3\times10^{-6}$  and $n_{\rm{c}}\approx 0.1$ photons [red solid curve in Fig.~\ref{splitting}d], which is several orders of magnitude smaller than for normal linear resonators~\cite{burnett2018noise,burnett2014evidence,brehm2017transmission,mcrae2020materials,kowsari2021fabrication,verjauw2021investigation,calusine2018analysis}. This can be explained by long-lived TLSs due to $n_{\rm{c}}\propto \sqrt{T_{1,\rm{TLS}}T_{2,\rm{TLS}}}$ \cite{gao2008physics}. Additionally, we also analyse the non-TLS loss where the loss could be possibly from both the chip design and chip modes [See supplementary].

Compared to the coherence time of the SNAIL-terminated resonator coupled to a waveguide near the Kerr-free point, the relaxation time here is significantly larger, indicating that the coherence time is limited by the pure dephasing. Assuming that the resonator coupled to the waveguide has the same internal relaxation time as the nonlinear resonator coupled to a qubit, according to $1/T_{\rm{s}}=1/(2T_1)+1/T_{\phi}$ with $T_1\approx 8~\mu s$ at $\kb{n}\approx 0$ from Fig.~\ref{splitting}d, we can infer the pure dephasing time $T_{\phi}\approx 12\unite{\mu s}$ and $2\unite{\mu s}$ at $\Phi_{\rm{ext}}=0$ and $\Phi_{\rm{ext}}=0.386~\Phi_{0}$, respectively. Especially, at $\Phi_{\rm{ext}}=0.386~\Phi_{0}$, the resonator becomes much more sensitive to the magnetic flux noise, leading to a shorter coherence time which could be increased by reducing the SNAIL parameter $\beta$ or adding more magnetic shielding to suppress the magnetic-flux noise or decrease the loop size.

%\begin{figure}[tbph]
%\centering
%\includegraphics[width=1\linewidth]{FIG7_cpg_snail.pdf}
%\caption{\textsf{{\bf{\textsf{Wigner distribution for the cubic phase state.}}}
%{\bf{a}}, Obtained from a master equation simulation with the
%Hamiltonian, single-photon loss rate $\kappa/2\pi=1/(2\pi T_1)\approx 20\unite{kHz}$ and the pure dephasing rate $\gamma_{\phi}/2\pi=1/(2\pi T_\phi)\approx 80\unite{kHz}$ by sequentially applying the squeezing and cubic phase gate to an initial vacuum state. The cubic phase state $|\xi,r\rangle=e^{i(\xi/2\sqrt{2}) (a^\dag+a)^3e^{(r/2)(a^{\dag2}-a^2)}}|0\rangle$, where $\xi$ is cubicity of the cubic phase gate applied,  $r$ the real squeezing parameter, and $|0\rangle$ the photon vacuum state (see~\cite{timo2020universal} for details).
%{\bf{b}}, Ideal cubic phase state with matched cubicity $\xi\approx0.1$ and squeezing
%$r\approx0.70ð (\approx6\unite{dB})$.
%}
%}
%\label{cpg_snail}
%\end{figure}

%\section{Discussion}
In this study, we developed an efficient method to characterize a SNAIL-terminated nonlinear resonator and observed well-resolved photon-number splitting up to nine photons, where the relaxation time can be up to $20\unite{\mu s}$. The parameters of our device are suitable for implementing universal gate sets for quantum computing~\cite{timo2020universal}. Compared to linear modes coupled to qubits, in our case, the Kerr effect from the ancilla qubit can be suppressed, leading to a better storage of bosonic modes where the Kerr coefficient makes the quantum state collapse \cite{kirchmair2013observation}.
%Based on the sample parameters, we numerically simulate the cubic-phase state [See methods], where the fidelity of the state is $95.56\%$ compared to the ideal case [Fig.~\ref{cpg_snail}].
Moreover, by taking the advantages of the nonlinearity of our resonator, we could also generate and store entangled photon modes and non-Gaussian states~\cite{chang2020observation,agust2020tripartite}.  The dispersive coupling between the qubit and the resonator could also enable us to make quantum state tomography very efficient. Finally, compared to 3D cavities, our chip-scale structure is more compact for integration and scalability required by the large-scale CV quantum processing in the future.

\section*{Supplementary}
{\bf{Experimental setup.}} The complete experimental setup is shown in Fig.~\ref{setup}. The sample, as shown in Fig.1 in the main text, is wire-bonded in a nonmagnetic oxygen-free cooper sample box, mounted to the 10\unite{mK} stage of a dilution refrigerator, and shielded by a $\mu$-metal can which is used for suppressing the static and fluctuating magnetic field. The signal to the qubit and the SNAIL-terminated resonator, with heavily attenuation, is combined to the charge line on the sample. A pulse is sent to the feedline on the sample to read out the state of the qubit with amplification using a traveling-wave parametric amplifier (TWPA) at the 10\unite{mK} stage, a high electron mobility transistor (HEMT) amplifier at 3\unite{K} and a room-temperature amplifier.

\begin{figure}[t!]
\includegraphics[width=1\linewidth]{FIGS1_setup1.pdf}
\caption{\textsf{{\bf{\textsf{Experimental setup}}.}
Complete wiring diagram and room temperature setup for the nonlinear resonator coupled to a qubit. The qubit operation and the pulsed displacement are made by analog upconversion of the pulsed generated by an  AWG (blue and orange). The dashed line in the blue box means that sometimes the microwave generator is connected to the combiner to pump the resonator continuously. The qubit readout (yellow) is controlled by the up- and down- conversion of a pulse generated by an AWG, and then digitized by an analog to digital converter (ADC).
}
}
\label{setup}
\end{figure}

{\bf{Sample Fabrication.}} We first clean a high-resistance ($10\unite{k\Omega}$) 2-inch silicon wafer by hydrofluoric acid with $2\%$ concentration. Then, we evaporate aluminum on top of the silicon substrate, followed by direct laser writing, and etching using wet chemistry to obtain all the sample details except the Josephson junction. The Josephson junctions for the qubit and the SNAIL are defined in a bi-layer resist stack using electron-beam lithography. Later, we deposit aluminum again by using a two-angle evaporation technique. In order to ensure a superconducting contact between the junctions and the rest of circuit, an argon ion mill is used to remove native aluminium oxide before the junction aluminium deposition. Finally, the wafer is diced into individual chips and cleaned properly using both wet and dry chemistry.

\begin{figure}[t!]
\includegraphics[width=1\linewidth]{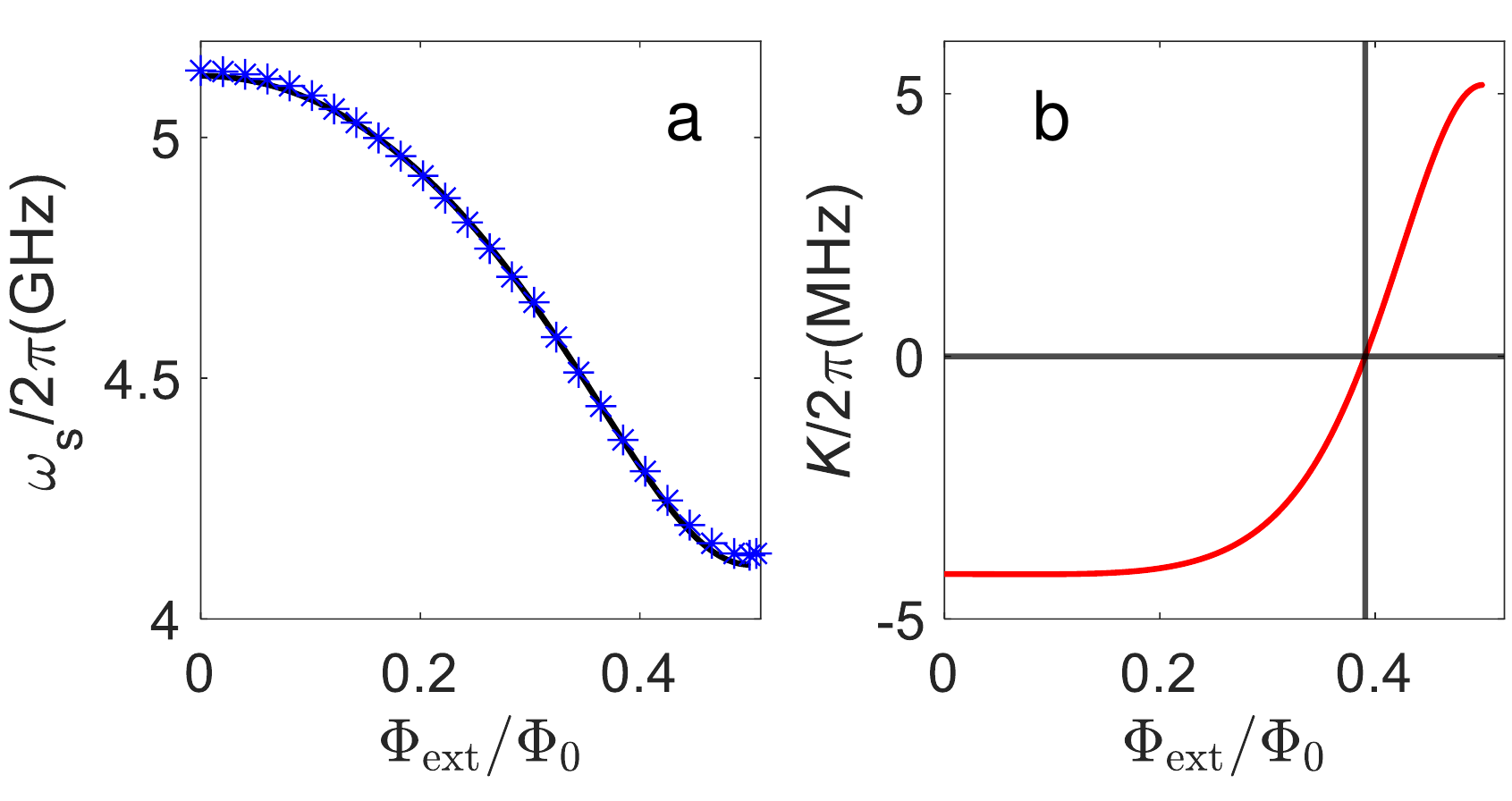}
\caption{\textsf{{\bf{\textsf{Frequency and Kerr strength for a SNAIL-terminated resonator coupled to a waveguide}}.}
{\bf{a}}, The frequency of the nonlinear resonator vs. the external magnetic flux through the SNAIL. Stars are the experimental data determined from the transmission coefficient of the vector analyzer network. The solid blue curve is calculated numerically.
{\bf{b}}, The Kerr-frequency shift per photon, $\sf{K}$, vs. the external magnetic flux through the SNAIL. The Kerr-free point occurs at $\Phi_{\rm{ext}}=0.39~\Phi_{0}$.
}
}
\label{baresnail}
\end{figure}

{\bf{Device parameters for the SNAIL-terminated resonator coupled to the waveguide.}} By sending a microwave probe to the SNAIL-terminated resonator coupled to the waveguide [the device for Table I in the main text], we measure the transmission coefficient of a SNAIL-terminated resonator to obtain the bare resonator frequency [blue stars in Fig.~\ref{baresnail}(a)]. Then, we get $\beta=0.095$ and $E_{\rm{J}}/(2\pi\hbar)=\frac{\hbar}{4e^2L_{\rm{J}}}\approx 830\unite{GHz}$ with the Josephson inductance $L_{\rm{J}}=600\unite{pH}$ by fitting the data to Eq.(3) in the main text [solid curve]. In Fig.~\ref{baresnail}(b), based on the values of $E_{\rm{J}}$ and $\beta$, we numerically calculate the Kerr coefficient $K$ due to the four-wave mixing coupling, which $K$ can be suppressed to zero at the flux point $\Phi_{\rm{ext}}/\Phi_0=0.39$.

\begin{figure}[t!]
\includegraphics[width=1\linewidth]{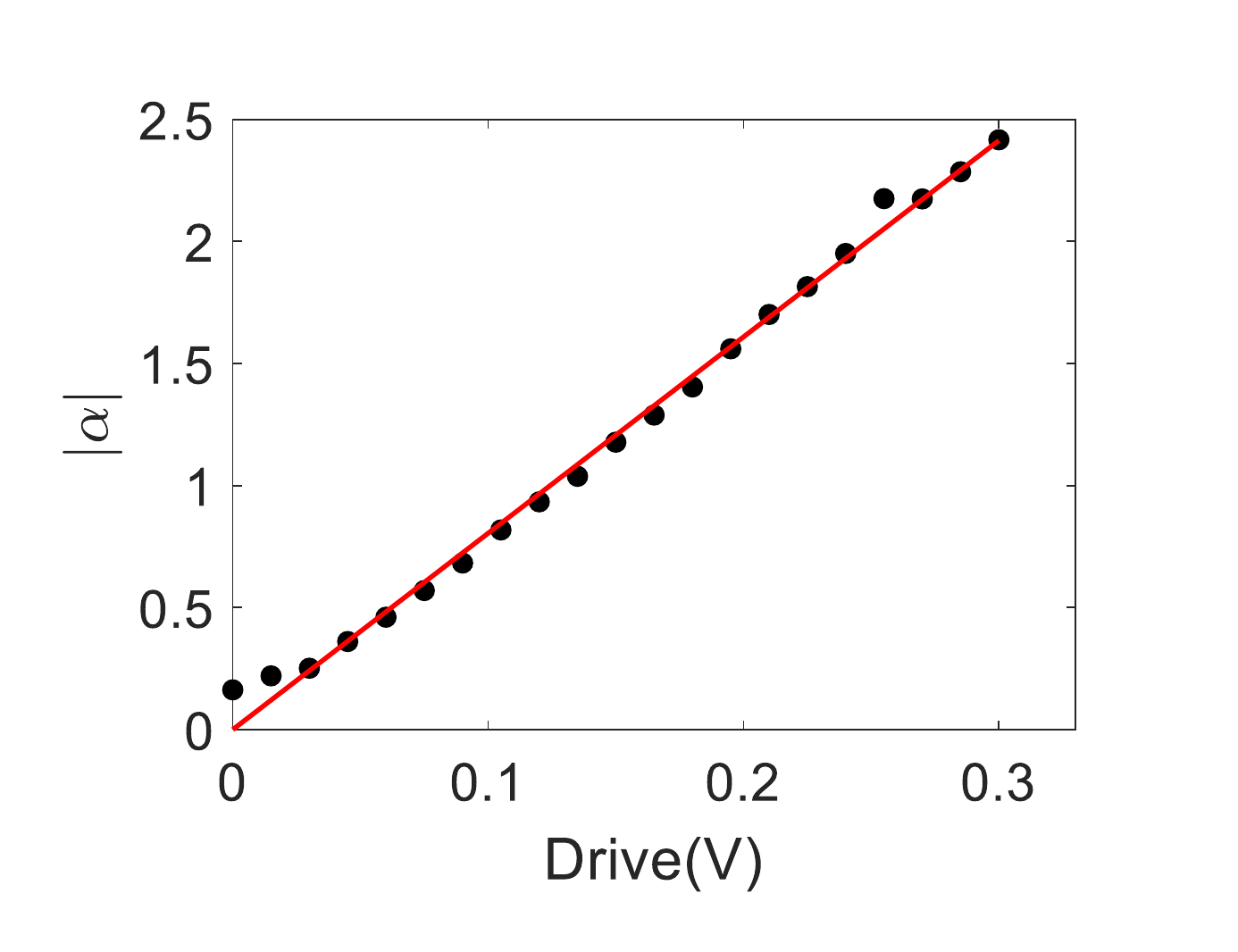}
\caption{\textsf{{\bf{\textsf{Pulse amplitude calibration}}.}
Calibration of the coherent displacement $\alpha$ with the voltage
output $A$, of the arbitrary-wavefunction generator at the room-temperature.
}
}
\label{calibration}
\end{figure}

\begin{figure}
\includegraphics[width=1\linewidth]{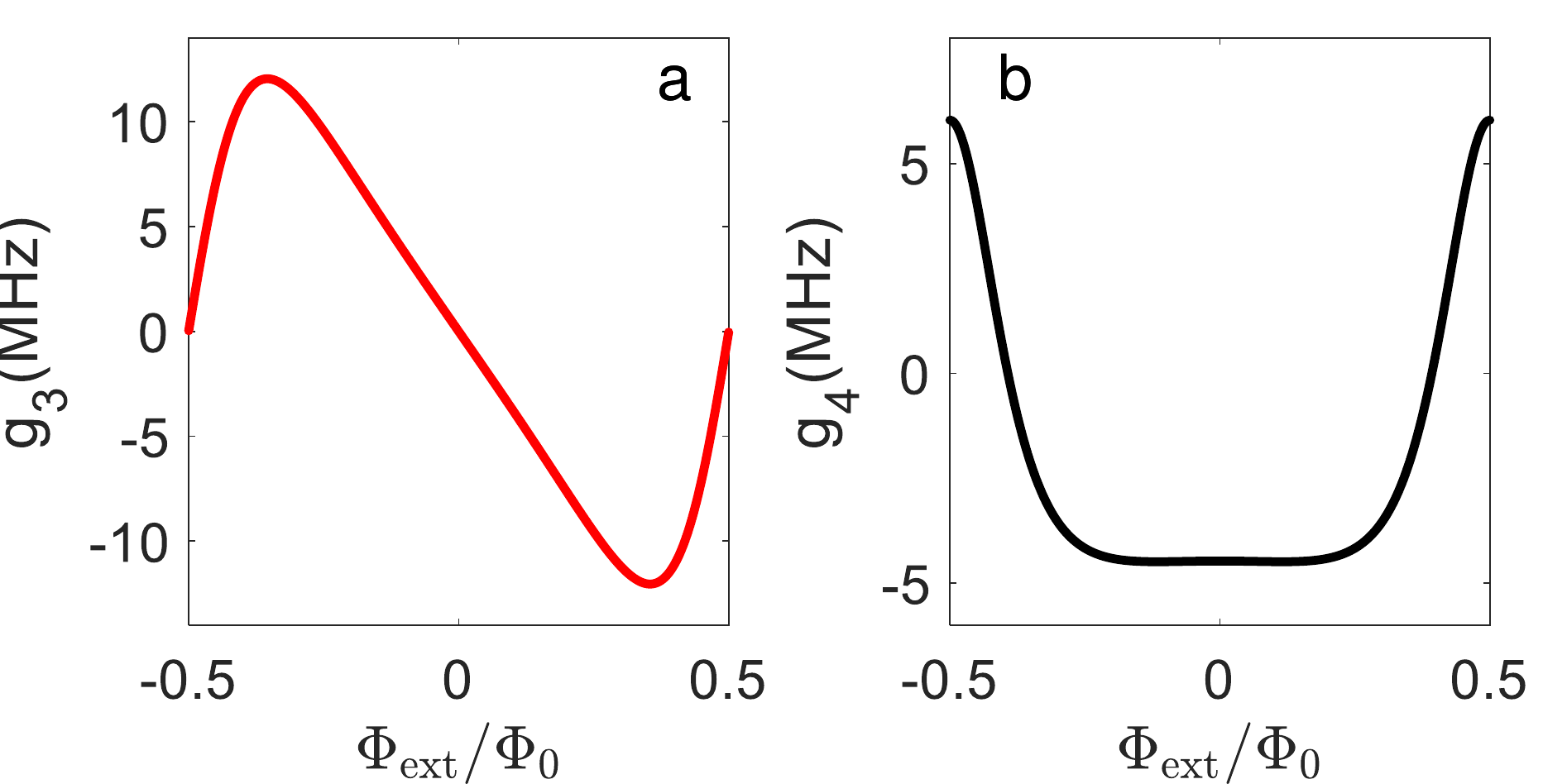}
\caption{\textsf{{\bf{\textsf{$g_3$ and $g_4$ numerical values for the SNAIL-terminated resonator coupled to a qubit}}.}
The numerical values of $g_3$ and $g_4$ for the SNAIL-terminated resonator coupled to a qubit.
}
}
\label{g3g4}
\end{figure}

\begin{figure}
\includegraphics[width=1\linewidth]{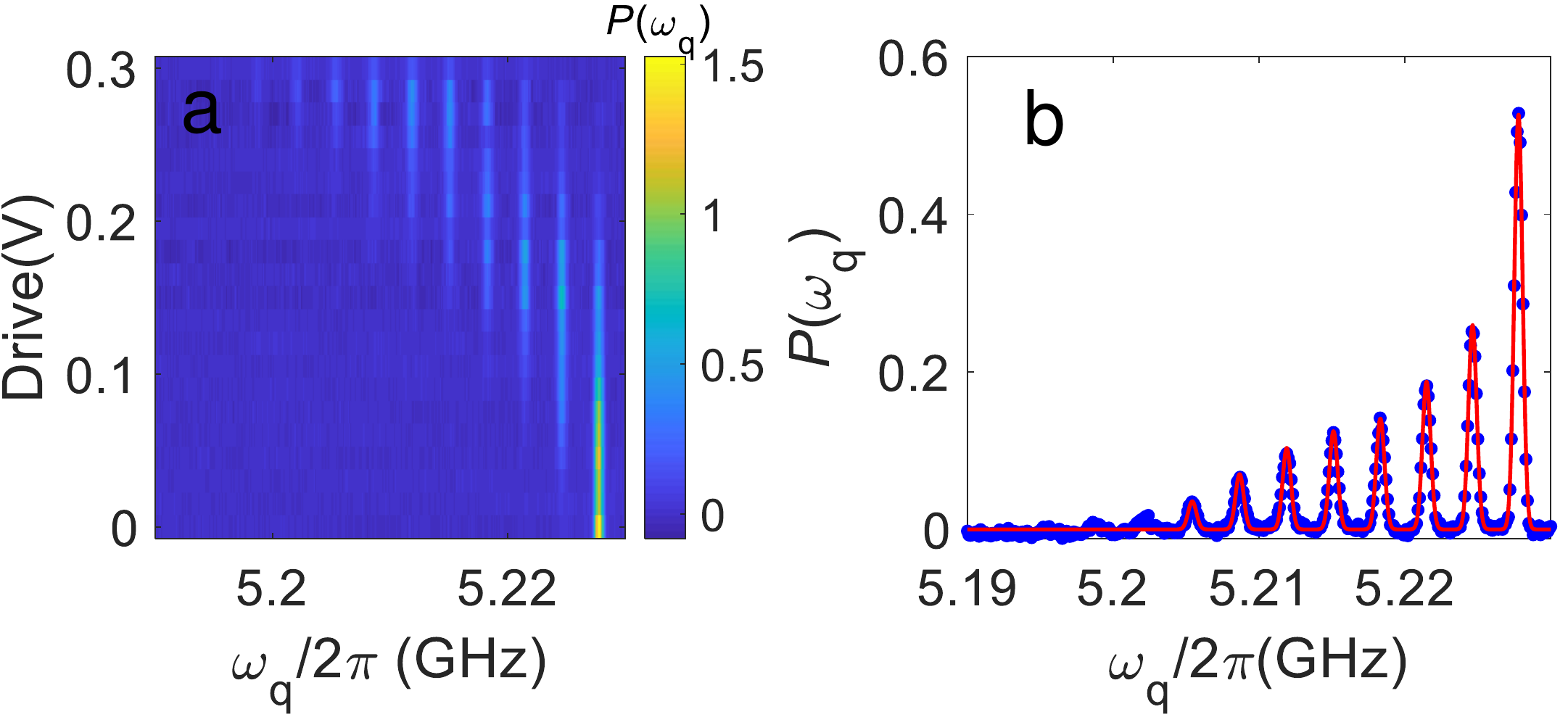}
\caption{\textsf{{\bf{\textsf{Qubit spectroscopy under different drive voltages}}.}
{\bf{a}}, The qubit-frequency distribution $P(\omega_{\rm{q}})=P(\omega_{\rm{q}})\times 10^6$ Vs. different drive voltages.
{\bf{b}}, The averaged qubit-frequency distribution.
}
}
\label{splittingS}
\end{figure}

{\bf{Pulse generation and calibration.}} The microwave control for the qubit and the nonlinear resonator is achieved by up-converting the inphase (I) and the quadrature (Q) components of a low frequency pulse generated by four AWG channels (two each for the qubit and the resonator). The qubit read-out pulse is up-converted to the sample and then down-converted after the amplification with the same local oscillator (yellow area). The pump pulse for the displacement is
generated by an IQ mixer with an input pulse from an
Arbitrary Waveform Generator (AWG) with amplitude. In Fig.~\ref{calibration}, we vary the amplitude of the pump pulse by changing the voltage amplitude $A$, and then obtain the corresponding displacement in the resonator according to the photon-number splitting as described in the main text. The data (black dots) is fitted to the equation $|\alpha|=kA$ with $k\approx 8$. This measurement is useful to calibrate the pump pulse with different amplitudes. We also notice that the residual thermal population with $A=0$ is $|\alpha|=0.16$ ($\kb{n}=|\alpha|^2\approx 0.03$ photons). The thermal and Poisson distributions are almost the same when the thermal photon number is small. In this case, we only take into account the single photon $P_1$ and the vacuum $P_0$ populations.Thus, we can calculate the corresponding thermal temperature as $T_{\rm{res}}=\hbar\omga{c}/\big[k_{\rm{B}}\ln{\big(1+\frac{1}{\kb{n}}}\big)\big]\approx 58\unite{mK}$.
{\bf{Kerr coefficient calculation.}} According the extracted parameters $\beta$ and $L_{\rm{J}}$ from Fig.~3a in the main text, we can calculate the coefficients $g_3$ and $g_4$ for the three-wave mixing and four-wave mixing terms, respectively [Fig.~\ref{g3g4}]. Then, we can infer the coupling strength for the Kerr coefficient $K$ according to $K=12(g_4-5g_3^2/\omga{s})$ [Fig.3b in the main text].
{\bf{Qubit spectroscopy.}} In order to increase the signal-to-noise ratio, here, we average the qubit spectroscopy with different driven voltages [Fig.~\ref{splittingS}a] to obtain Fig.~\ref{splittingS}b. Afterwards, each peak corresponding to a photon number is fitted to a Gaussian function to extract the dispersive shift as shown in Fig.5b in the main text.

\begin{figure}[t!]
\includegraphics[width=1\linewidth]{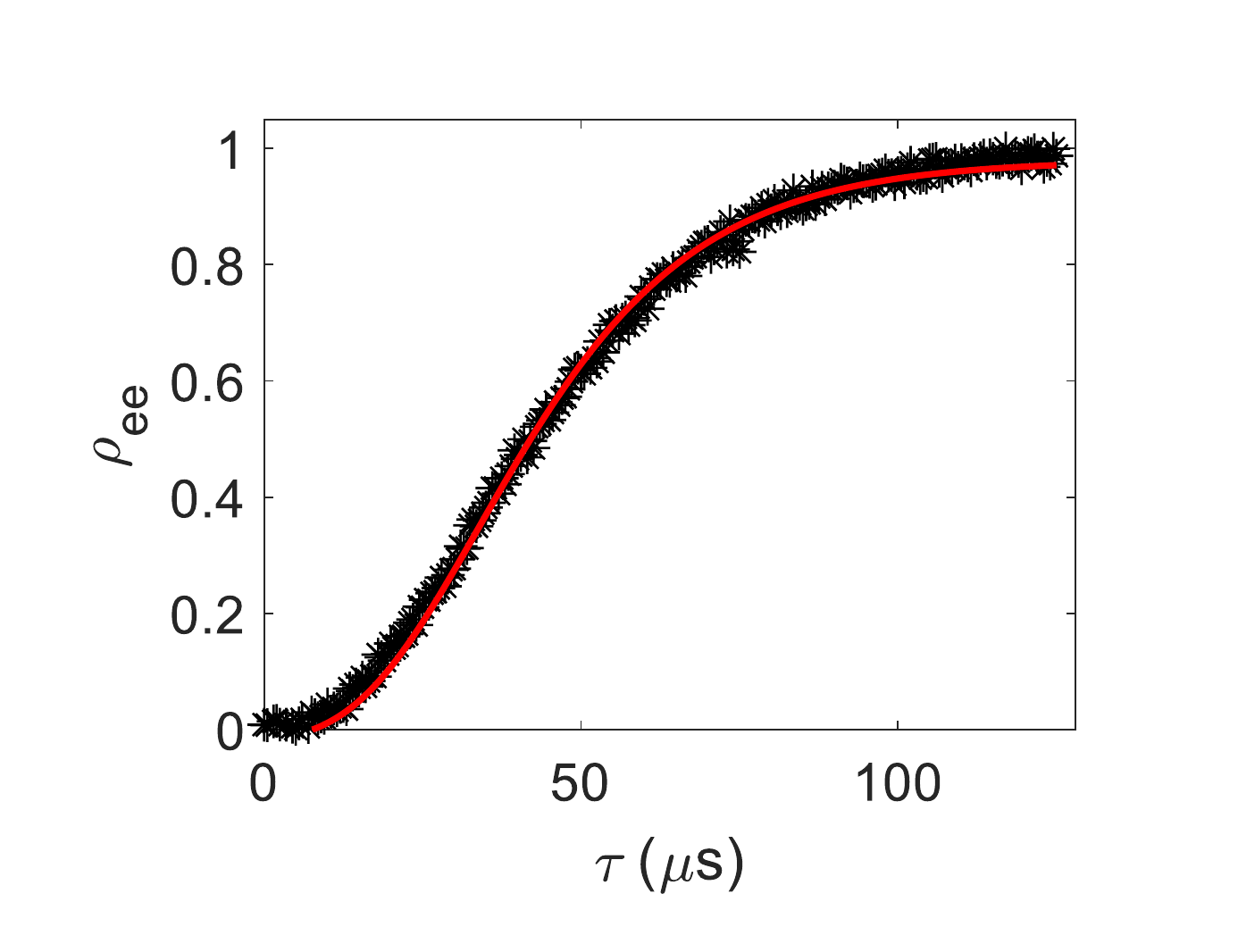}
\caption{\textsf{{\bf{\textsf{Extract the value of $T_1$ from the qubit population}}.}
The qubit population $\rho_{\rm{ee}}$ vs. the waiting time $\tau$ with $\kb{n}=5.8$.
}
}
\label{T1}
\end{figure}

{\bf{The lifetime $T_1$ of the resonator}}. In Fig.~\ref{T1}, The data trace [black stars] is from the top trace in Fig.5c in the main text with the average photon number $\kb{n}=5.8$. Here, we show an example to obtain the value of $T_1$ from Fig.5c in the main text.
The qubit excitation probability is $\rho_{\rm{ee}}=1-\rm{exp}[-|\alpha(0)|^2{\rm{exp}(-\tau/T_1)}]$. By fitting the rising curve [Fig.~\ref{T1}] with the calibrated $|\alpha(0)|=\sqrt{5.8}$ $(\kb{n}=5.8)$ , we extract $T_1=19.2\pm0.2\unite{\mu s}$, corresponding to a quality factor of $Q_1=\omga{c}T_1\approx5.18\times10^5$ which is close to the $Q_1$ value for a coplanar linear resonator in the few-photon limit~\cite{kowsari2021fabrication,burnett2018noise}.
In order to improve $T_1$ further in the future, we also analyse the non-TLS loss. The non-TLS loss limits the lifetime to be $T_{\rm{other}}=1/(\delta_{\rm{other}}\omga{c})\approx 28\unite{\mu s}$ comparable to our qubit lifetime $T_{1,\rm{q}}\approx 20\unite{\mu s}$. In our sample, the nonlinear resonator couples to the qubit, charge line and a flux line where they contribute to the non-TLS loss as $\gamma_{\rm{q}}/2\pi=(g_0/\Delta_0)^2\frac{1}{T_{1,\rm{q}}}\approx170\unite{Hz}$, $\gamma_{\rm{c}}/2\pi\approx2.07\unite{kHz}$ and $\gamma_{\rm{f}}/2\pi\approx477\unite{Hz}$, respectively (the values of $\gamma_{\rm{c}}$ and $\gamma_{\rm{f}}$ are from the simulation). In total, the loss from the chip design is $\gamma_{\rm{total}}=\gamma_{\rm{q}}+\gamma_{\rm{c}}+\gamma_{\rm{f}}\approx2\pi\times2.717\unite{kHz}$ corresponding to $T_{\rm{total}}=1/\gamma_{\rm{total}}\approx 59{\unite{\mu s}}\approx 2 T_{\rm{other}}$´, meaning that other unknown losses such as the chip modes are similiar to the loss from the chip design.

%{\bf{System parameters.}} We summarize all the parameters for the whole system in Table

%++++++++++++++++++++++++++++++++++++++++++++++++++++++++++++++++++++++++++++++++++++++++
% Yes methods is after discussion, for some reason they do that in Nature Physics
%+++++++++++++++++++++++++++++++++++++++++++++++++++++++++++++++++++

\section*{Acknowledgements}
The authors acknowledge the use of the Nano fabrication Laboratory (NFL) at Chalmers. We also acknowledge IARPA and Lincoln Labs for providing the TWPA used in this experiment.
We wish to express our gratitude to Lars J\"{o}nsson and Xiaoliang He for help and we appreciate the fruitful discussions with Axel Eriksson, Simone Gasparinetti and Zhirong Lin. This work was supported by the Knut and Alice Wallenberg Foundation via the Wallenberg Center for Quantum Technology (WACQT) and by the Swedish Research Council.
%++++++++++++++++++++++++++++++++++++++++++++++++++++++++++++++++++++++++++++++++++++++++
\section*{AUTHOR CONTRIBUTIONS}
Y.L. planned the project. Y.L. performed the measurements with help from M.K. and J.Y.Y. Y.L. designed and fabricated the sample. T.H. and F.Q. helped to make the numerical calculation. Y.L. wrote the manuscript with input from all the authors. Y.L. analyzed the data. P.D. supervised this project.

\section*{COMPETING INTERESTS}
The authors declare no competing interests.

\section*{REFEREENCES}
%\bibliographystyle{naturemag}

%\bibliography{library}

\end{document}